\documentclass[letterpaper, 11pt]{article}

\usepackage[top=1in, bottom=1in, left=1in, right=1in]{geometry}
\usepackage{enumitem}
\setlist{listparindent=\parindent}
\setlist{parsep=\parskip}

\usepackage{xcolor}
\usepackage[colorlinks, allcolors=blue]{hyperref}
\usepackage{newtxtext}
\usepackage{newtxmath}

\usepackage[style=authoryear-comp,
            maxcitenames=1,
            maxbibnames=5,
            uniquelist=false,
            sortcites=false,
            natbib]{biblatex}
\bibliography{refs.bib}

\usepackage{amsmath}
\usepackage{amssymb}
\usepackage{mathtools}
\usepackage{graphicx}
\usepackage{wrapfig}
\usepackage{hyperref}
\usepackage{biblatex}
\usepackage{todonotes}
\usepackage{authblk}

\newcommand{\bigO}{\mathcal{O}}

\renewcommand{\vec}[1]{\mathbf{#1}}
\newcommand{\uvec}[1]{\vec{e}_{#1}}
\newcommand{\units}[1]{\; \mathrm{#1}}

\newcommand{\pdiff}[1]{\partial_{#1}}

\newcommand{\pderiv}[2]{\frac{\pdiff{} #1}{\pdiff{} #2}}

\newcommand{\grad}[1]{\nabla #1}
\newcommand{\dive}[1]{\nabla \cdot #1}

\newcommand{\lap}[1]{\nabla^2 #1}

\renewcommand{\eqref}[1]{Eq.~(\ref{#1})}

\newcommand{\figref}[1]{Fig.~\ref{#1}}
\newcommand{\tabref}[1]{Table~\ref{#1}}

\newcommand{\Benard}[0]{B\'{e}nard}
\newcommand{\Danger}{\mathfrak{D}}

\begin{document}

\title{The ``Sphered Cube'': A New Method for the Solution of Partial Differential Equations in Cubical Geometry}

\author[1]{Keaton~J.~Burns}
\author[2,3]{Daniel~Lecoanet}
\author[4]{Geoffrey~M.~Vasil}
\author[5]{Jeffrey~S.~Oishi}
\author[6]{Benjamin~P.~Brown}

\affil[1]{Center for Computational Astrophysics, Flatiron Institute, NY 10010, USA}
\affil[2]{Princeton Center for Theoretical Science, Princeton, NJ 08544, USA}
\affil[3]{Princeton University Department of Astrophysical Sciences, Princeton, NJ 08544, USA}
\affil[4]{University of Sydney School of Mathematics and Statistics, Sydney, NSW 2006, Australia}
\affil[5]{Bates College Department of Physics and Astronomy, Lewiston, ME 04240, USA}
\affil[6]{University of Colorado Laboratory for Atmospheric and Space Physics and Department of Astrophysical and Planetary Sciences, Boulder, CO 80309, USA}

\date{}

\maketitle

\begin{abstract}

A new gridding technique for the solution of partial differential equations in cubical geometry is presented.
The method is based on volume penalization, allowing for the imposition of a cubical geometry inside of its circumscribing sphere.
By choosing to embed the cube inside of the sphere, one obtains a discretization that is free of any sharp edges or corners.
Taking full advantage of the simple geometry of the sphere, spectral bases based on spin-weighted spherical harmonics and Jacobi polynomials, which properly capture the regularity of scalar, vector and tensor components in spherical coordinates, can be applied to obtain moderately efficient and accurate numerical solutions of partial differential equations in the cube.
This technique demonstrates the advantages of these bases over other methods for solving PDEs in spherical coordinates.
We present results for a test case of incompressible hydrodynamics in cubical geometry: Rayleigh-\Benard{} convection with fully Dirichlet boundary conditions.
Analysis of the simulations provides what is, to our knowledge, the first result on the scaling of the heat flux with the thermal forcing for this type of convection in a cube in a sphere.

\end{abstract}

\section{Introduction}

Cubes are ubiquitous in everyday life. 
In addition to environmental and industrial applications \citep[e.g.,][and references within]{hunt1991,partridge2017}, cubes are known to form in nature due to cubical crystal structures\footnote{\url{https://io9.gizmodo.com/this-chunk-of-fools-gold-naturally-formed-in-these-cube-1689943134}}. 
However, modelling partial differential equations (PDEs) in cubical geometries is plagued with numerical difficulties. 
Previous work has used a range of numerical techniques to solve equations in cubes (or the 2D equivalent, squares). 
E.g., \citet{basak2006} used a finite element scheme and \citet{dixit2006} used a lattice Boltzmann method for low resolution 2D simulations of natural convection in a square. 
The most straightforward technique is simple finite difference methods \citep[e.g.,][]{kuznetsov2010}. 
In each case, previous methods have used a grid aligned with the cube axes, which may cause features to artificially align with these axes.
Furthermore, care must be taken at the edges and corners, where one must ensure boundary conditions satisfy consistency conditions.
To avoid these issues, we introduce a new algorithm for the simulation of PDEs in cubical geometries.

Our approach is inspired by previous methods which have successfully simulated PDEs in spherical geometries by remapping the surface of the sphere to a cube.
\citet{sadourny1972} introduced this technique, dubbed the ``cubed sphere'' by \citet{ronchi1996}, as a way to avoid the coordinate singularities at the poles of the sphere, which complicate the discretization of spherical vector and tensor components.
The cubed sphere has proved highly successful as a starting point for finite-volume discretizations of the sphere.
The method is widely used in the ocean and atmospheric science communities, and is utilized by several major general circulation models \citep{adcroft2004,putnam2007}.

Here we introduce a novel method---the ``sphered cube''---for solving general PDEs in cube geometry, by first embedding the cube into a sphere, and then using the volume penalization to apply Dirichlet boundary conditions on the faces of the cube. 
This builds on the recent work of \citet{vasil2019} \& \citet{lecoanet2019}, which develops elegant methods for solving arbitrary tensorial PDEs in a full sphere using global basis functions which correctly account for the behavior of these quantities near the coordinate singularities.
Volume penalization can then be used to damp variables (e.g., velocities, temperature, density, magnetic field, etc.) toward a prescribed boundary values (which in general can be functions of space and time). 

To illustrate the power and versatility of the ``sphered cube,'' we include novel simulation results of Rayleigh-\Benard{} convection in cube geometry. 
Using volume penalization, we impose that there is no-slip and zero deviation from the background linear temperature gradient at the boundaries of the cube.
This allows for non-zero heat flux and thermal boundary layers along the sides of the cube. 
These simulations represent (as far as we know) the first investigation of Rayleigh-\Benard{} convection in this geometry using these boundary conditions.

\section{The sphered cube}

\subsection{Volume penalization}

\begin{figure}
    \centering
    \includegraphics[width=0.5\textwidth]{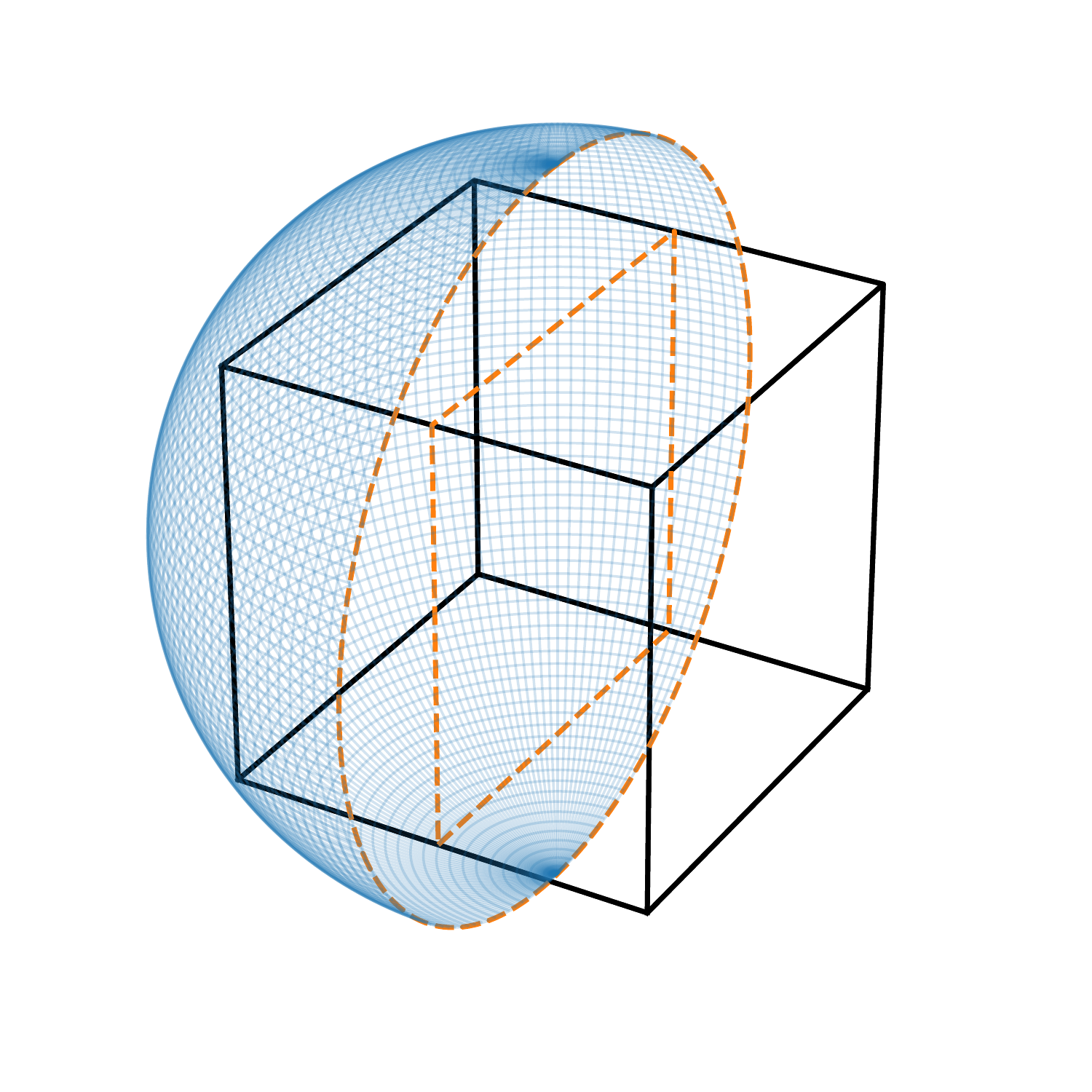}
    \caption{The sphere and the cube constituting the sphered cube.  
    A PDE in the cube is extended the its circumscribing sphere, with volume penalization damping the solution in the extension (outlined in orange) to a reference solution to mimic Dirichlet boundary conditions on the cube.}
    \label{fig:sphered_cube}
\end{figure}

The sphered cube technique solves PDEs posed in a cube of side-length $L$ by extending them to the circumscribing sphere of radius $R = \sqrt{3} L /2$.
The volume penalization technique is then used to approximately enforce Dirichlet boundary conditions on the PDE variables at the now internal surfaces corresponding to the faces of the cube, known as the \emph{fictitious boundary}.
See \figref{fig:sphered_cube} for an illustration of the geometry.

The volume penalty technique involves adding a masked damping term to prognostic equations in a PDE to drive the corresponding variables towards reference solutions within the masked regions.
Specifically, an evolution equation for a quantity $C$ is augmented with a damping term as
\begin{equation}
    \pderiv{C}{t} = ... + \frac{\Gamma(\vec{x})}{\tau}\left(C - C_\mathrm{ref}(\vec{x})\right)
\end{equation}
\noindent Here $\Gamma(\vec{x})$ is a spatially-varying \emph{mask function} which tends to $0$ in the original domain and $1$ in the extension, $\tau$ is the imposed damping timescale, and $C_\mathrm{ref}(\vec{x})$ is the reference solution for the exterior which is designed to match the desired Dirichlet boundary conditions at the fictitious boundary.

Many volume penalization implementations utilize discontinuous masks that are aligned with the fictitious boundary.
However, for quantities diffusing with some diffusivity $\kappa$, a boundary layer of size $\bigO(\sqrt{\kappa \tau})$ forms in the vicinity of the fictitious boundary, leading to $\bigO(\tau^{1/2})$ errors in the interior solution.
\cite{Hester:2019vk} derive and demonstrate that shifting or smoothing the mask by an amount $\bigO(\sqrt{\kappa \tau})$ can eliminate this error and improve the convergence of the method to $\bigO(\tau)$.

Our spectral algorithm requires the mask function to be smooth.
We therefore choose our mask function to be
\begin{equation}
    \Gamma(\vec{x}) = \frac{1}{2}\left[1 + \tanh\left(\frac{\max(|x|,|y|,|z|)-L/2}{\delta}\right)\right].
\end{equation}
\noindent where the Cartesian coordinates $(x,y,z)$ have their origin at the center of the sphere and the smoothing scale of the mask is controlled by $\delta \ll L$.
The mask is close to zero in the cube defined by $x,y,z \in (-L/2,L/2)^3$ and close to unity outside.
Although here we only consider a stationary mask outlining a cube, the same technique can be used to model a wide variety of shapes, including time-varying domains.

\subsection{Rayleigh-\Benard{} convection}

To illustrate the utility of the method, we consider the problem of incompressible Rayleigh-\Benard{} convection in a cube.
The Boussinesq equations for an incompressible fluid with thermal buoyancy consist of conservation equations for mass, momentum, and heat:
\begin{equation}
    \dive{\vec{u}} = 0
\end{equation}
\begin{equation}
    \pderiv{\vec{u}}{t} + \vec{u} \cdot \grad{\vec{u}} = - \grad{p} + \nu \lap{\vec{u}} - \alpha \vec{g} T
\end{equation}
\begin{equation}
    \pderiv{T}{t} + \vec{u} \cdot \grad{T} = \kappa \lap{T}
\end{equation}
\noindent Here $\vec{u}$ is the fluid velocity, $p$ is the pressure (divided by the constant fluid density), $\nu$ is the kinematic viscosity, $\alpha$ is the thermal expansivity, $\vec{g} = - g \uvec{z}$ is the gravity vector, $T$ is the fluid temperature, and $\kappa$ is the thermal diffusivity.

Standard Rayleigh-\Benard{} convection considers Boussinesq fluid between two horizontal plates, separated by a distance $L$, with the bottom plate held at a temperature $\Delta T$ higher than the top.
The sidewalls are taken to be perfectly insulated and there is no slip of the fluid around the entire boundary.
We will consider an alternative formulation where a decreasing linear background temperature is enforced at all points on the boundary:
\begin{equation}
    T_\mathrm{boundary} = - \frac{z}{L} \Delta T
\end{equation}
\noindent The non-dimensional form of these equations using the box size $L$ as the length scale, $\Delta T$ as the temperature scale, and the thermal diffusive time $\tau_\kappa = L^2 / \kappa$ as the time scale are:
\begin{equation}
    \dive{\vec{u}} = 0
\end{equation}
\begin{equation}
    \pderiv{\vec{u}}{t} + \vec{u} \cdot \grad{\vec{u}} = - \grad{p} + \mathrm{Pr} \lap{\vec{u}} + \mathrm{Pr} \mathrm{Ra} T \uvec{z}
\end{equation}
\begin{equation}
    \pderiv{T}{t} + \vec{u} \cdot \grad{T} = \lap{T}
\end{equation}
\begin{equation}
    T_\mathrm{boundary} = - z \quad \textrm{and} \quad \vec{u}_\mathrm{boundary} = 0
\end{equation}
\noindent with the non-dimensional control parameters being the Prandtl number $\mathrm{Pr} = \nu / \kappa$ and the Rayleigh number and $\mathrm{Ra} = \alpha g \Delta T L^3 \nu \kappa = \tau_\nu \tau_\kappa / \tau_\mathrm{ff}^2$ where $\tau_\mathrm{ff} = \sqrt{L / \alpha g \Delta T}$ is the free-fall time.

We will further consider the equations in terms of the perturbation temperature $T' = T - T_\mathrm{ref} = T + z$ and impose the boundary conditions via volume penalization with a dimensional damping time $\tau$ and masking function $\Gamma(\vec{x})$, resulting in the final system:
\begin{equation}\label{eqn:incompressible}
    \dive{\vec{u}} = 0
\end{equation}
\begin{equation}
    \pderiv{\vec{u}}{t} + \vec{u} \cdot \grad{\vec{u}} = - \grad{p'} + \mathrm{Pr} \lap{\vec{u}} + \mathrm{Pr} \mathrm{Ra} T' \uvec{z} - \epsilon \Gamma \vec{u}
\end{equation}
\begin{equation}\label{eqn:temperature}
    \pderiv{T'}{t} + \vec{u} \cdot \grad{T'} - \vec{u} \cdot \uvec{z} = \lap{T'} - \epsilon \Gamma T'
\end{equation}
\noindent with the additional non-dimensional control parameter $\epsilon = \tau_\kappa / \tau$ controlling the strength of the volume penalization.
Homogeneous Dirichlet boundary conditions on $\vec{u}$ and $T'$ are additionally applied at the numerical domain boundary, i.e. the surface of the sphere.

Although this paper focuses on the specific fluid dynamics example of Rayleigh-\Benard{} convection in a sphere, the algorithm could just as easily solve other PDEs in this geometry, e.g., heat conduction in a rectangular, Detroit-style pizza.

\section{A variation of the GSZ polynomial method}

We solve Eqs.\ \ref{eqn:incompressible}--\ref{eqn:temperature} in a sphere of non-dimensional radius $\sqrt{3}/2$ using a spectral method based on the Generalized Spherical Zernike (GSZ) polynomials introduced in \citet{vasil2019}, and implemented using the Dedalus\footnote{More information at \url{http://dedalus-project.org}.} code \citep{burns2016} in \citet{lecoanet2019}. 
The calculation uses spherical coordinates, $(r,\theta,\phi)$. 
Each variable is represented as a series expansion in spectral bases in each direction. 
In the $\phi$ direction we use Fourier series, in the $\theta$ direction we use spin-weighted spherical harmonics, and in the $r$ direction we use GSZ polynomials.

The spin-weighted spherical harmonics automatically satisfy the necessary regularity conditions at the poles for arbitrary tensorial quantities. 
Scalar quantities are expanded using spin-zero spherical harmonics. 
The components of a vector can be recombined into ``spin components,'' having spins of $-1$, $0$, and $+1$, each of which we represent with the appropriate spin-weighted spherical harmonic basis. 
This system extends in a straightforward way to general tensor quantities (e.g., Reynolds stresses, etc.). 
This expansion allows for smooth flows across the poles while properly capturing the discontinuous behavior of the angular components of vectors and tensors there.

Similarly, for the radial basis we use radially weighted Jacobi polynomials (GSZ polynomials) which satisfy the appropriate regularity conditions at $r=0$. 
This representation requires recombinations of the different spin-components of a tensor, where the specific recombination depends on the spherical harmonic degree $\ell$. 
The details of this recombination and the associated GSZ polynomials are discussed \textit{ad nauseam} in \citet{vasil2019}.

Tensorial derivative operators (the gradient, curl, and Laplacian) act in a simple and sparse way on GSZ polynomials. 
Derivatives act on GSZ polynomials by multiplying by a simple factor and changing the basis, similar to differentiation of a sine series which multiplies by the wavenumber and converts to a cosine series. 
Nonlinear products are most efficiently calculated in physical space. 
This requires transforming the data to physical space, for which we use the FFT in the $\phi$ direction, and matrix multiplication transforms in the $\theta$ and $r$ directions. 
An exhausting verification of the code is detailed in \citet{lecoanet2019}.

As discussed in \citet{lecoanet2019}, we solve equations~\ref{eqn:incompressible}--\ref{eqn:temperature} by jointly evolving the state vector
\begin{equation}
    X = \left[V^-, V^0, V^+, p', T'\right]^T,
\end{equation}
\noindent where $V^-$, $V^0$, and $V^+$ are the three regularity components of the velocity vector in spectral space. 
We solve the discretized PDE system
\begin{equation}
    M \cdot \pderiv{X}{t} + L \cdot X = N(X). 
\end{equation}
\noindent where $M$ and $L$ are the sparse block matrices formed by discretizing the PDE operators in the chosen spectral bases. 
The terms on the left hand side of the equals sign are timestepped implicitly, and the terms on the right hand side of the equals sign are timestepped explicitly.
The matrices (defined below) are $\ell$-dependent, but do not couple different $\ell$ or $m$ modes together (where $\ell$ and $m$ are the spherical harmonic and azimuthal degree). 
Thus, we can solve for the radial structure of each $\ell$ and $m$ mode separately. 
The matrices are

\begin{equation}
    M \ = \ \left[
\begin{array}{ccccc}
 C_{1,\ell-1}C_{0,\ell-1} & 0 & 0 & 0 & 0 \\
 0 & C_{1,\ell}C_{0,\ell} & 0 & 0 & 0 \\
 0 & 0 & C_{1,\ell+1}C_{0,\ell+1} & 0 & 0 \\
 0 & 0 & 0 & 0 & 0 \\
 0 & 0 & 0 & 0 & C_{1,\ell}C_{0,\ell}
\end{array}
\right],
\end{equation}

\begin{equation}
\small
L \ = \ \left[
\begin{array}{ccccc}
 -\mathrm{Pr}R^{-2}D_{1,\ell}^-D_{0,\ell-1}^+ & 0 & 0 & R^{-1}\xi_\ell^-C_{1,\ell-1}D_{0,\ell}^- & 0 \\
 0 & -\mathrm{Pr}R^{-2}D_{1,\ell+1}^-D_{0,\ell}^+ & 0 & 0 & 0 \\
 0 & 0 & - \mathrm{Pr}R^{-2}D_{1,\ell}^+D_{0,\ell+1}^- & R^{-1}\xi_{\ell}^+C_{1,\ell+1}D_{0,\ell}^+ & 0 \\
 \xi_{\ell}^-D_{0,\ell-1}^+ & 0 & \xi_{\ell}^+D_{0,\ell+1}^- & 0 & 0 \\
 0 & 0 & 0 & 0 & - R^{-2}D_{1,\ell+1}^-D_{0,\ell}^+
\end{array}
\right].
\end{equation}

\noindent The $C$ are basis conversion matrices, $D$ are derivative matrices, and $\xi^\pm$ are $\ell$-dependent prefactors, all defined in \citet{lecoanet2019}. 
$R=\sqrt{3}/2$ is the radius of the sphere, which must be included in the matrices because the $D$ are derivative operators for a sphere with unit radius. 
The explicitly timestepped terms are

\begin{equation}
    N(X) \ = \ \left[
\begin{array}{c}
 C_{1,\ell-1}C_{0,\ell-1}\left(-\vec{u}\vec{\cdot}\vec{\nabla}\vec{u} + \textrm{Pr}\textrm{Ra}\, T'\, \vec{e}_z - \epsilon\Gamma \vec{u}\right)^- \\
 C_{1,\ell}C_{0,\ell}\left(-\vec{u}\vec{\cdot}\vec{\nabla}\vec{u} + \textrm{Pr}\textrm{Ra}\, T'\, \vec{e}_z- \epsilon\Gamma \vec{u}\right)^0 \\
 C_{1,\ell+1}C_{0,\ell+1}\left(-\vec{u}\vec{\cdot}\vec{\nabla}\vec{u} + \textrm{Pr}\textrm{Ra}\, T'\, \vec{e}_z- \epsilon\Gamma \vec{u} \right)^+ \\
 0 \\
 C_{1,\ell}C_{0,\ell}\left(\vec{u}\cdot\vec{e}_z - \vec{u}\vec{\cdot}\vec{\nabla} T' - \epsilon\Gamma T'\right)
\end{array}
\right].
\end{equation}

\noindent Note that the volume penalization terms and the $\vec{u}\cdot\vec{e}_z$ terms are linear in the problem variables, but have $\theta$ and/or $\phi$ dependent coefficients and thus couple different $\ell$ and $m$ modes together. 
Thus, we evaluate these terms in physical space rather than in spectral space. 
This is similar to the treatment of the Coriolis effect in rotating convection in \citet{lecoanet2019}, and this technique can be used for any general term that couples $\ell$ and $m$ modes.
The only meaningful degree of freedom when $\ell=0$ is the temperature field. 
In this case we only keep the bottom right entry to the $M$ matrix (zeroing out the rest), and keep the bottom right entry of the $L$ matrix, but replace the upper left 4x4 block with the identity. 
The $N(X)$ function is taken to be zero for $\ell=0$.

Although we use volume penalization outside the cube, we still must apply boundary conditions at the sphere's surface to regularize the system. 
We enforce no-slip and no temperature perturbation at the surface (matching the conditions applied at the surface of the cube via volume penalization) using a generalized $\tau$ method \citep{lanczos38,vasil2019}.

\section{Results on single processor computers}

An important metric for the computational performance of the sphered cube method is its geometrical \textit{efficiency factor}, that is, the ratio of the volume of the cubic domain of interest compared to the total volume being discretized within in the sphere,
\begin{equation}
    \frac{V_c}{V_s} < \frac{3}{\sqrt{2} \pi} \simeq 0.675.
\end{equation}
\noindent Thus, calculations in the sphered cube require approximately fifty percent more work than volume penalization in a triply periodic domain only slightly larger than the cube itself. 
Furthermore, our parallelization strategy is not evenly load balanced \citet{lecoanet2019}, resulting in some cores doing approximately twice as much work as would be required with perfect load balancing.
The current performance of the sphered cube implementation in Dedalus therefore renders the utilization of single processor computers inadvisable.

\section{The ``sphered cube'' on massively parallel architectures}

\subsection{Parameter selection}

We solve the discretized system described above for several Rayleigh numbers beyond the convective threshold.
These calculations require high resolution but are automatically parallelized using MPI via the Dedalus framework and can thus be executed with relative efficiency on thousands of processes.
Our parameters are chosen as follows:

\begin{itemize}

\item The maximum spherical harmonic degree $L_\mathrm{max}$ and maximum radial polynomial degree $N_\mathrm{max}$ are taken to be equal: $L_\mathrm{max} = N_\mathrm{max} = N$ and we apply $3/2$-dealiasing to each dimension when transforming between grid values and coefficients.

\item The Prandtl number is taken to be unity.

\item Estimating the large-scale turbulent velocity to be roughly $U \approx L / \tau_\mathrm{ff}$ allows us to estimate the Reynolds number as $\mathrm{Re} = L U / \nu \approx \tau_\nu / \tau_\mathrm{ff}$ and $\mathrm{Ra} \approx \mathrm{Re}^2 \mathrm{Pr}$.
We choose a target Reynolds number for our simulations as that from Kolmogorov's theory for turbulence with an injection scale of $L$ and a dissipation scale equal to the grid scale $\Delta x = R / N$, namely $\mathrm{Re} \approx (L / \Delta x)^{4/3}$.
We then take $\mathrm{Ra} = \Danger \mathrm{Pr} N^{8/3}$, where $\Danger$ is the \emph{danger factor} of the simulation. 
We found a nice rational danger factor to be $\Danger = 256/9$, which provides a good balance between computational efficiency and numerical stability.

\item The volume penalization timescale is chosen so that the corresponding lengthscale is comparable to the grid/dissipation scale: $\sqrt{\nu \tau} \approx \Delta x$ or specifically $\epsilon = \frac{32}{3} \mathrm{Pr} N^2$.
The mask width is also chosen to match the grid/dissipation scale: $\delta = \Delta x$.
We note that these choices place our simulations in the ``intermediate'' damping region described in \cite{Hester:2019vk}, where volume penalization achieves $\bigO(\tau)$ convergence irrespective of the details of the mask.
These parameters were chosen because they result in resolution requirements equal to those imposed by the turbulent flow itself, and therefore provide the highest-accuracy volume-penalized simulation that can be achieved with no computational overhead.

\item The simulation timestep is limited to $\epsilon^{-1}/2$ for stability in the explicit integration of the volume penalization terms.
The SBDF2 timestepper \citep{wang2008} is used with an adaptive timestep based on a CFL criterion and a CFL safety factor (which should more appropriately be termed the ``CFL danger factor'' as increasing this factor reduces accuracy and stability) of $0.4$.
The CFL criterion is calculated using the radial grid-crossing timescale for the radial flow but the timescale $R / L_\mathrm{max} u$ for the angular flow, since the method is robust to fast advection across the closely clustered angular grid near the origin and polar axis.
The CFL limit is estimated to be $\Delta x / U \approx \tau_\mathrm{ff} / N \approx \mathrm{Re}^{-7/4} \tau_\nu$, which is smaller than the restriction from the damping.

\end{itemize}

\subsection{Simulation results}

\begin{figure}[h]
    \centering
    \includegraphics[width=1.0\textwidth]{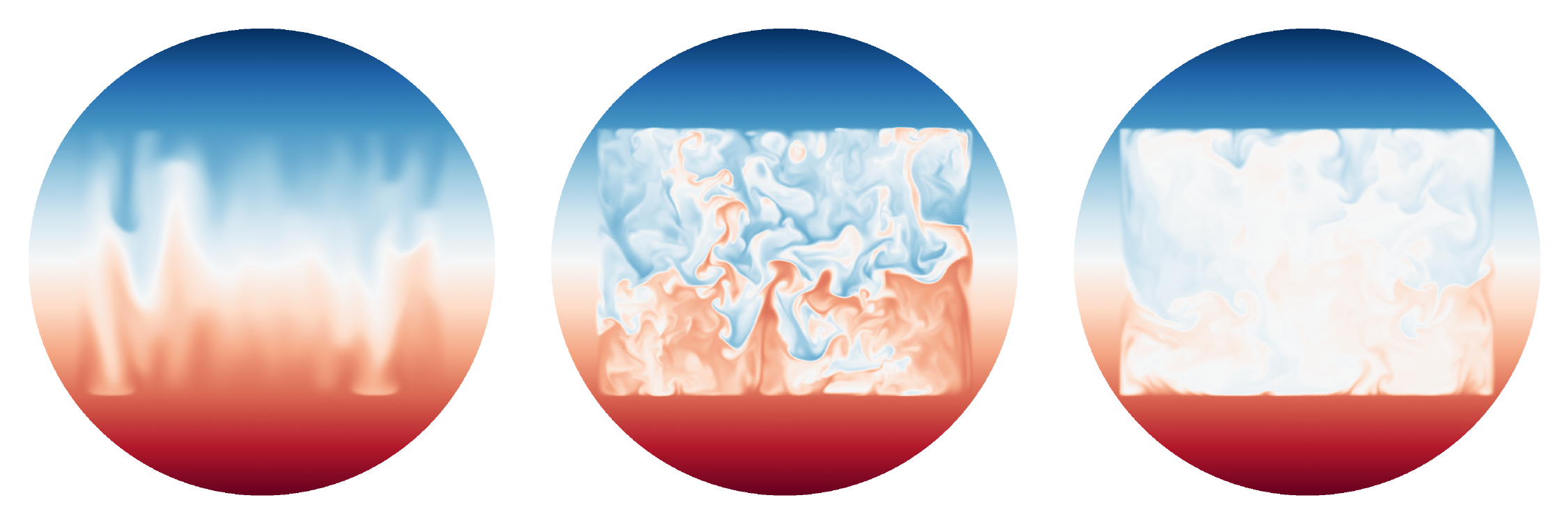}
    \caption{Temporal evolution of a vertical slice of the total temperature field of the $N=383$ simulation.
    The slice is taken along the $x-y$ diagonal of the inscribed cube, so gravity is vertically downwards.
    From left to right, the images show the simulation late in the linear instability, at the transition to turbulence, and in the steady convective state.
    Videos available at \url{http://data.dedalus-project.org/sphered_cube/}.}
    \label{fig:triptych}
\end{figure}

\begin{table}[h]
\centering
\begin{tabular}{cccc|c}
   N &  Pr &            Ra &       $\epsilon$ &         Nu \\
\hline
 127 &   1 &  1.159e+07 &  1.720e+05 &  17.81 \\
 255 &   1 &  7.438e+07 &  6.936e+05 &  34.67 \\
 383 &   1 &  2.201e+08 &  1.565e+06 &  52.25 \\
\end{tabular}
\caption{Simulation input parameters (resolution, Prandtl number, Rayleigh number, damping strength), and diagnosed parameters (Nusselt number).}
\label{tab:sims}
\end{table}

We perform simulations at various resolutions and fixed danger factors and Prandtl numbers (see \tabref{tab:sims}).
Trials with larger danger factors were found to be numerically unstable and exhibited spectral ringing in the convective steady state.
We initialize the velocity to zero, and seed the convection instability with cell-by-cell, Gaussian distributed random temperature perturbations with amplitude $10^{-3}$. 
The initial phase of the calculation tracks the linear growth. 
Figure~\ref{fig:triptych} shows a slice of the evolution of the total temperature field through the linear instability, the transition to turbulence, and into the steady convective state.

\begin{figure}[h]
    \centering
    \includegraphics[width=0.6\textwidth]{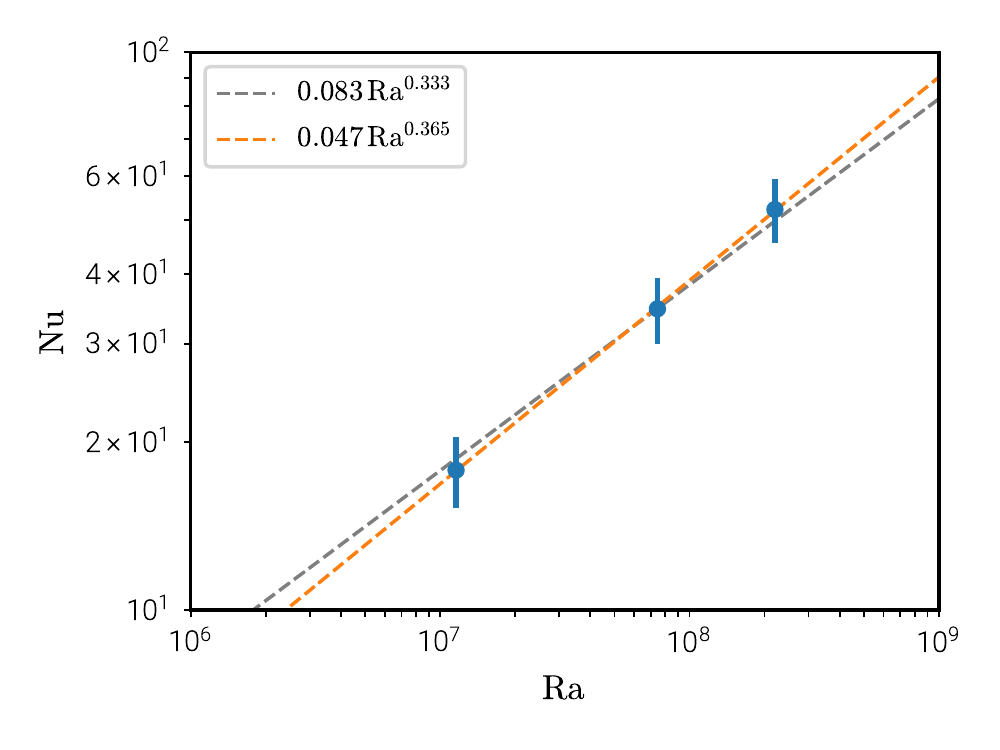}
    \caption{Measured Nusselt number as a function of Rayleigh number for three simulations (blue points).
    The vertical bars indicate the standard deviation of the Nusselt number in time in the saturated convective state.
    The best-fitting power law is given in orange.
    The fit with the exponent fixed to $1/3$ is given in grey.}
    \label{fig:nusselt}
\end{figure}

The standard quantity of interest in Rayleigh-\Benard{} convection is the heat-transfer enhancement due to convection, relative to the conduction of the background state, which in our non-dimensionalization is
\begin{equation}
    \mathrm{Nu} = \langle \vec{u} \cdot \uvec{z} T' - \grad{T} \rangle
\end{equation}
\noindent where the brackets denote a horizontal and temporal average in the convective state.
In the standard Rayleigh-\Benard{} setup with insulating sidewalls, this quantity is independent of height.
In our formulation, this quantity varies in height due to lateral heat fluxes at the wall.
We therefore choose to measure the Nusselt number at the midplane ($z = 0$ or $\theta = \pi/2$).
Furthermore, we neglect mean-state conductive contribution, which is small.
The midplane Nusselt number, averaged over many free-fall times in the convective steady state, is plotted as a function of the Rayleigh number in \figref{fig:nusselt}.
Since three points more than suffice to draw a line, we perform a least-squares fit to estimate the power-law relation between these quantities, which is found to be $\mathrm{Nu} \approx 0.047 \mathrm{Nu}^{0.365}$.
It is unknown if this scaling represents the ultimate asymptotic behavior of Dirichlet Rayleigh-\Benard{} convection in the sphered cube.

An alternative metric of the impact of these results is the carbon footprint of running our simulations.
The largest simulation $N=383$ required approximately $50,000$ core-hours to simulate for $100$ free-fall times.
Taking a per-core energy consumption of $0.02 \units{kW}$ for a supercomputing facility such as NASA Pleiades\footnote{\url{https://www.top500.org/system/177259}} and a carbon generation rate of $215 \units{kg/MWh}$ for California\footnote{\url{https://www.eia.gov/electricity/state/california/}}, we estimate that this simulation is responsible for $215 \units{kg}$ of carbon emission.

\section{Conclusions}

The sphered cube is a method for solving partial differential equations in cubical domains by extending the equations to the circumscribing sphere and imposing Dirichlet boundary conditions via volume penalization.
This method utilizes recently developed spherical bases which enable sparse tensor calculus operations while exactly incorporating the varying regularities of scalar, vector, and tensor components in spherical coordinates.
We have used the method to simulate Rayleigh-\Benard{} convection in the turbulent regime.

This exercise demonstrates the flexibility and robustness of this spectral method for simulating equations inside the sphere.
First, the method easily accommodates broad ranges of PDEs including the Navier-Stokes equations in the Boussinesq approximation with volume penalization.
Second, the method is robust to strong flows passing through the origin and along the polar axis.
Finally the method allows for the stable integration of such flows with timesteps respecting a CFL criterion based on the total spherical harmonic order rather than the local spacing of the angular grid, which becomes vanishingly small near the polar axis.

In the future, we plan to utilize these spherical bases to simulate rough and evolving boundaries in the sphere via the phase-field method.
Potential applications to other compactly-supported fixed domains also abound, including the possibility of developing a spectral-element method based on coupling simulations in many sphered tetrahedra.

\printbibliography

\end{document}